**Angle and Polarization Selective Spontaneous Emission in Dye-doped Metal/Insulator/Metal Nanocavities**

*Vincenzo Caligiuri*, Giulia Biffi, Milan Palei, Beatriz Martin-Garcia, Renuka Devi Pothuraju, Yann Bretonnière, and Roman Krahne**

((Optional Dedication))

Dr. V. Caligiuri, G. Biffi, Dr. M. Palei, Dr. B. Martin-Garcia, P.R. Devi, Dr. R. Krahne,
Istituto Italiano di Tecnologia, Via Morego 30, 16163 Genova, Italy
E-mail: roman.krahne@iit.it; vincenzo.caligiuri@iit.it

Dr. V. Caligiuri
Dipartimento di Fisica, Università della Calabria, 87036 Rende, Italy

G. Biffi, R. D. Pothuraju,
Dipartimento di Chimica e Chimica Industriale, Università degli Studi di Genova, Via Dodecaneso, 31, 16146 Genova, Italy

Dr. Y. Bretonnière,
Université de Lyon, Ecole Normale Superieure de Lyon, CNRS UMR 5182, Université Claude Bernard Lyon 1, Laboratoire de Chimie, F69342, Lyon, France

Keywords: Epsilon-Near-Zero, Polarized Spontaneous Emission, Push-Pull Chromophores, metal-insulator-metal cavities

Directing and polarizing the emission of a fluorophore is of fundamental importance in the perspective of novel photonic sources based on emerging nano-emitter technologies. These two tasks are usually accomplished by a sophisticated and demanding structuring of the optical environment in which the emitter is immersed, or by non-trivial chemical engineering of its geometry and/or band structure. In this paper, the wavelength and polarization selective spontaneous emission from a dye-embedded in a Metal/Insulator/Metal (d-MIM) nanocavity is demonstrated. A push-pull chromophore with large Stokes shift is embedded in a MIM cavity whose resonances are tuned with the spectral emission band of the chromophore. Angular and polarization resolved spectroscopy experiments reveal that the radiated field is reshaped according to the angular dispersion of the nanocavity, and that its spectrum manifests two bands with different polarization corresponding to the p- and s-polarized resonances of the cavity. The d-MIM cavities are a highly versatile system for polarization and wavelength division multiplexing applications at the nanoscale, as well as for near-field focused emission and nanolenses.

## 1. Introduction

Achieving control on the directionality and polarization of the emission of a fluorophore has important implications in modern light source technology based on emerging gain materials.[1–4] Usually, the spectral, spatial, and polarization control of the emission of a gain material is obtained by embedding it into either an optically structured environment like diffraction gratings,[5–7] waveguides,[8,9] photonic crystals,[10–14] and quasi-crystals,[15] or in a randomly organized one like a liquid crystal.[16–18] Recently, *hyperbolic metamaterials* (HMMs) have emerged as an elegant and versatile alternative to more common optical platforms. Thanks to their hyperbolic dispersion, HMMs show a very high local density of photonic states that allow to gain control on both the directionality and the photophysical properties of fluorophores embedded into them, or placed in their close proximity.[19–29] However, the noticeable light manipulation capabilities of HMMs are limited to near field interactions and providing far field control requires additional outcoupling techniques such as nanoscale plasmonic gratings. Together with beam routing, controlling the polarization state of the emitted radiation is of importance for optical information coding and multiplexing applications. Such a task is usually accomplished by acting directly on the gain material. Polarized emission has been demonstrated in strongly elongated colloidal quantum rods,[30] Ge films,[31] and suitably structured perovskite nanocrystals.[32,33] A promising alternative to such sophisticated techniques is constituted by MIM nanocavities.[34–36] The resonant behavior occurring in subwavelength MIM cavities has been subject of study for several years.[37] Modes with purely plasmonic nature, mixed photonic/plasmonic and purely photonic properties have been theoretically predicted,[38–41] and experimentally studied.[42–45] Thanks to their straightforward fabrication, Metal/Insulator/Metal (MIM) cavities of various geometries have been investigated, with applications in a plethora of fields, from superabsorption to high-resolution color printing.[46–54] More recently, MIM cavity resonances have been attributed to resonant tunneling Epsilon-Near-Zero (ENZ) modes that can be excited without the need of momentum matching techniques.[55,56] MIM structures are highly versatile to modify the photophysical properties of fluorophores. For example, by tuning the resonances of a MIM double cavity to the optical transition bands of a green emitting perovskite film, the Purcell effect together with spontaneous emission and quantum yield enhancement has been demonstrated.[57]

In this work, we demonstrate how to modulate the emission of a fluorophore by embedding it into an Epsilon-Near-Zero MIM nanocavity. We chose a novel fluorophore consisting of a push-pull chromophore ((E)-4-(2-(9-ethyl-9H-carbazol-3-yl)vinyl)-5,5'-dimethyl-2-oxo-2,5-dihydrofuran-3-carbonitrile, from now on GE133) with a large Stokes shift as emitting material due to its broadband photoluminescence, and the potential design opportunities in terms of refractive index, emission wavelength, *etc*.. We demonstrate both theoretically and experimentally that the angular dispersion of the d-MIM cavity induces angle selective light emission from the cavity, *i.e.* different wavelengths are outcoupled at different angles. In particular, we found that the spectrum of the light radiated out of the d-MIM system at a given angle is characterized by two modes that stem from the p- and s-polarization resonances of the cavity. Here the high-energy mode is s-polarized and the low-energy one is p-polarized.

## 2. Results and Discussion

**Figure 1a** shows the ellipsometrically measured and Scattering Matrix Method (SMM) simulated angular dependence of the p- and s-polarized modes of a 40/168/100 nm Ag/$Al_2O_3$/Ag MIM cavity. Such a system represents a MIM superabsorber,[43,55] and its resonances with p- and s-polarization blue-shift with increasing angle of incidence. The Q-factor associated to these modes obtained by SMM is plotted in Figure S1 of the Supporting

Information. The simulated optical response of a dye-doped superabsorber with the same geometry is shown in Figure 1b-e, where the fluorophore is modelled as a point dipole oriented horizontally in x-direction, embedded in the insulator layer. Here we plot the profile of the p-polarized norm of the electric field ($\sqrt{E_x^2 + E_y^2}$) calculated *via* finite element method (COMSOL) simulations (see Figure S2 in Section 2 of the SI for calculations of the s-polarised modes based on a magnetic dipole). The photons emitted by the fluorophore inside the cavity are outcoupled with the same angular dispersion that we extracted from the absorbance of the bare cavity, with shorter wavelengths emitted at larger angles. For example, a monochromatic p-polarized wave at λ = 635 nm is both absorbed (Figure 1a) and outcoupled (Figure 1c) at 62°. Here the ripple-like substructure in the light cones results from interference of the angular emission with the radiated field of the epsilon-near-zero (ENZ) modes, which propagate inside the cavity and radiate at normal incidence to the surface. At wavelengths between 680 nm and 690 nm, the angular dispersion is relatively flat, which leads to focusing of the radiated light in a narrow cone around the normal to the surface. For smaller wavelengths around 600 nm (for p-polarized light) the emission is almost at grazing angles to the surface of the MIM, resulting in a wave propagating on the surface of the MIM structure.

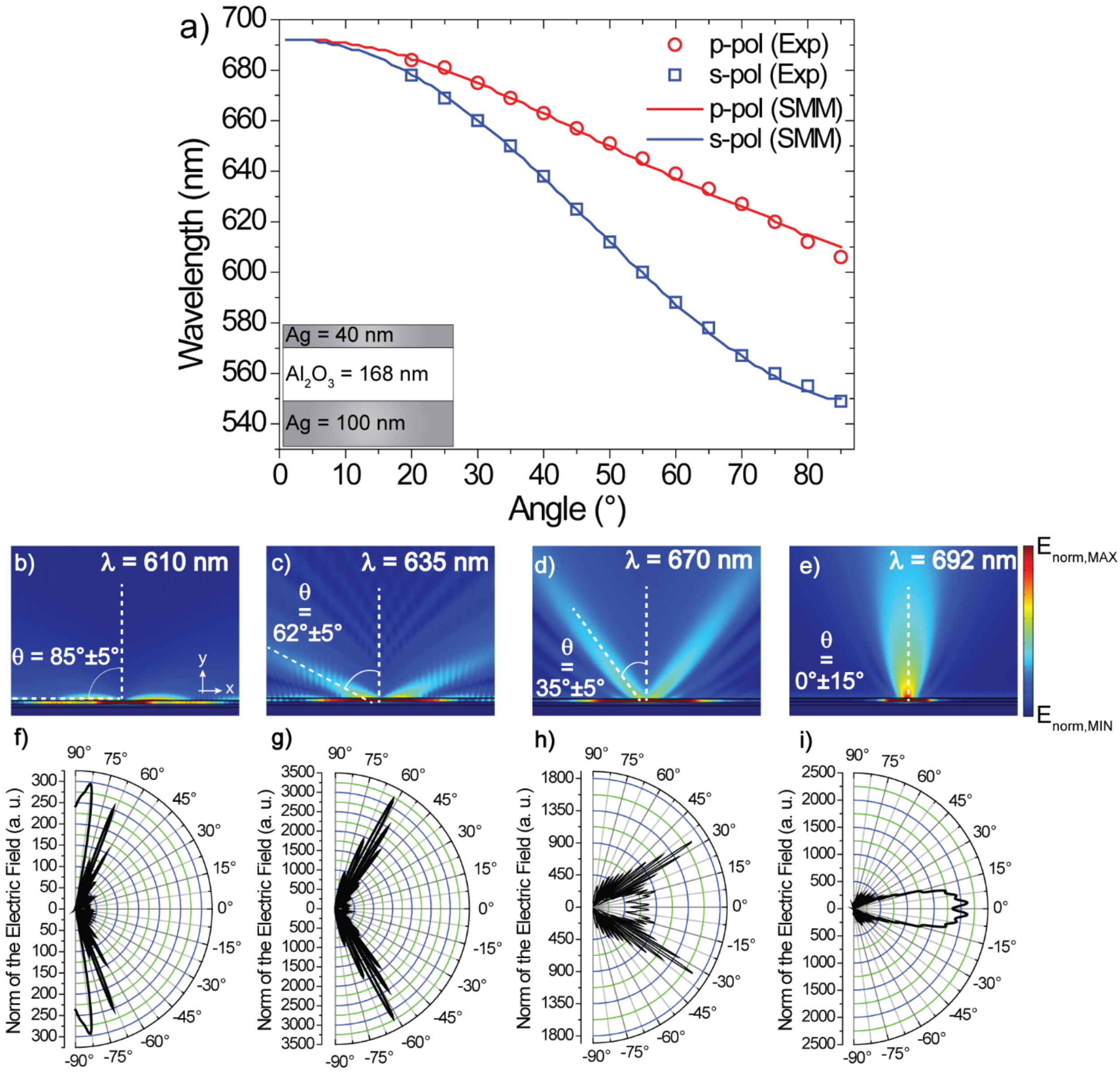

**Figure 1.** (a) Peak positions of the ellipsometrically measured p- and s-polarized resonances in absorbance (A = 1-T-R) measured by ellipsometry (red circles and blue squares, respectively) simulated by SMM (red and blue solid lines, respectively) for different angles of incidence. The structure is a 40/168/100 nm Ag/$Al_2O_3$ MIM superabsorber. (b-e) Norm of the electric field ($\sqrt{E_x^2 + E_y^2}$) in a similar MIM superabsorber calculated *via* COMSOL simulations for a horizontally oriented (x-direction) point dipole placed in the dielectric core of the MIM at (b) 600 nm, (c) 635 nm, (d) 670 nm, and (e) 692 nm. (f-i) Polar diagrams relative to panels (b-e).

The far-field emission of the dipole for the four wavelengths discussed in Figure 1 is plotted in the polar diagrams of Figure 1f-i, where an excellent agreement between the calculated emission wavelength and the angular distribution of measured absorbance maxima is found (similar polar diagrams for the s-polarised modes are shown in Figure S2).

A dye-doped MIM nanocavity with a 55 nm thick layer of an organic dye (see inset of **Figure 2b**) in the insulator layer has been fabricated as reported in the Materials and Methods section. The dye is a particular push-pull chromophore (GE133) that was synthesized according as to a published protocol,[58] with broadband emission and large transition dipole moment, which enables the fabrication of homogeneous films by spin-coating. One important feature of the GE-133 chromophore is that it has a high refractive index that is very close to the one of the $Al_2O_3$ layer in the MIM structure, which prevents abrupt changes of the optical properties at their interface. The p- and s-polarization resonances of the d-MIM have been designed in order to match the emission spectrum of the GE133 (see Figure S3 of the SI). In Figure 2a, the experimental p- and s- polarization reflectance spectra of the d-MIM are shown. The absorbance of GE133 falls in the range from 400-500 nm, highlighted by the grey color, leading to a broad dip in reflectance. The mode labeled with a capital "F" is the well-known Ferrell-Berreman mode, the natural ENZ mode of Ag.[59–62]. The two modes labeled "A-s" and "A-p" are, respectively, the s- and p-polarized antisymmetric modes of the d-MIM (see Figure S4 in Section 4 of the SI for simulations of the cross-section profiles of these modes). In Figure 2b, the photoluminescence (PL) of the GE133 in an open cavity configuration (without the top metal layer, as sketched in the left inset of Figure 2b) is compared with that of a "closed" d-MIM (see sketch in the right inset of Figure 2b). For the open cavity, the spectral shape of the PL is similar to the spontaneous emission of the pristine GE133. Time-resolved PL measurements shown in Figure S5 reveal a slightly faster PL decay when the chromophore layer is positioned directly on top of the Ag backreflector, which can be explained by partial quenching of the chromophore emission due to the metal layer. This effect can be mitigated by an additional layer of $Al_2O_3$ in between the Ag backreflector and the GE-133 layer, as demonstrated in Section 5 of the SI. For the closed d-MIM cavity, the shape of the PL spectrum is substantially different. Two peaks are present, of which the high-energy one corresponds to the s-polarization d-MIM symmetric mode and the low-energy one to the p-polarized symmetric mode. Here, the portion of the emission spectrum of the GE133 that is in resonance with the cavity is enhanced, while the emission at off resonance wavelengths is reduced. As discussed in detail in ref. [55], a MIM cavity can be seen as a homogenized layer with an effective permittivity, and this treatment can be extended to the d-MIM system. However, the dye in the MIM cavity introduces a gain component whose effect in the effective permittivity of the d-MIM results in a much more complex dispersion with multiple zero-crossings of both the real and imaginary part, as shown in Figure S6 of the SI.

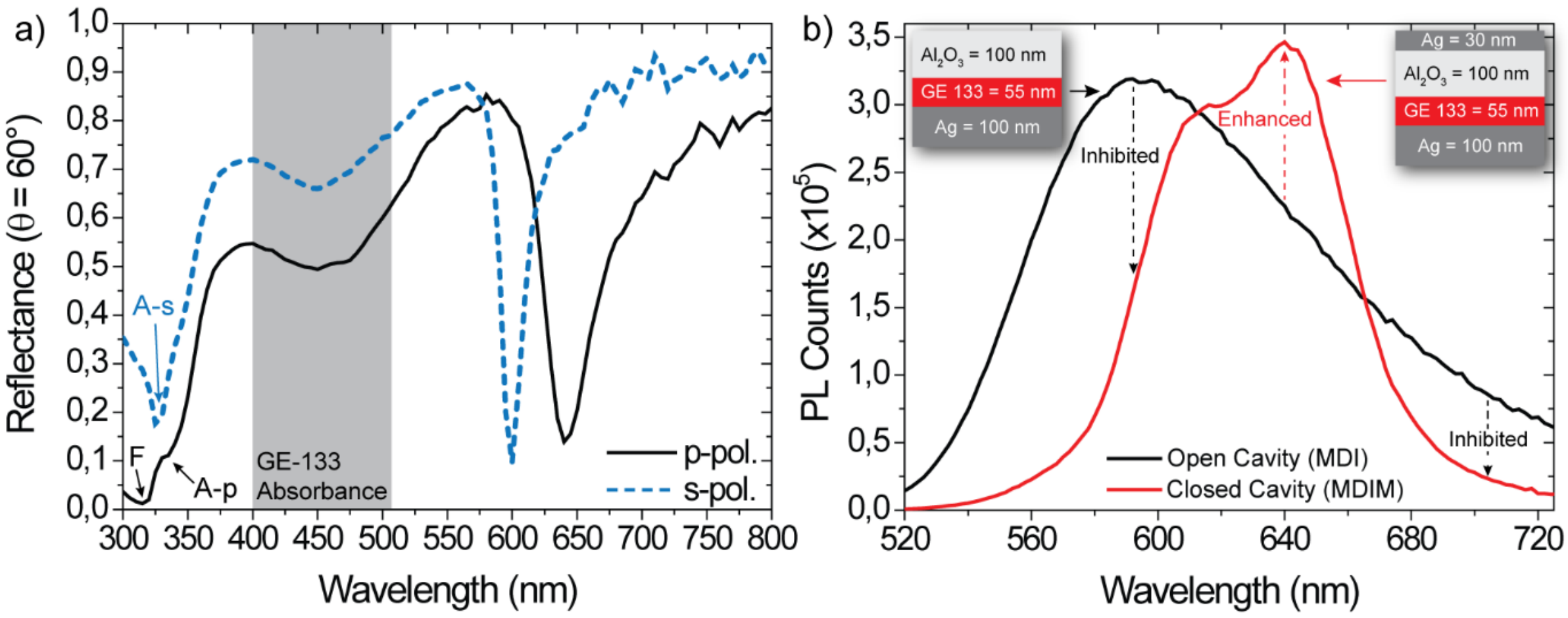

**Figure 2.** (a) P-polarized and s-polarized reflectance spectra of the d-MIM system collected at an angle of 60° to the surface normal. The p- and s-polarized symmetric modes are present at ~ 640 nm and ~ 600 nm. At shorter wavelength the Ferrell-Berreman mode (F), and the antisymmetric p- and s-polarized modes of the d-MIM cavity labeled “A-p” and “A-s”, can be identified. The absorbance range of the GE133 is highlighted in gray. (b) Photoluminescence of the GE133 fluorophore acquired at an angle of 60° for an open (black) and closed (red) cavity. Spectral regions of enhanced and reduced emission are indicated by red and black arrows, respectively.

In order to investigate the angular emission properties of the d-MIM, we used a setup as illustrated in **Figure 3a**. A pulsed laser pump beam at 405 nm excites the chromophore at an angle $\theta_{in}$, and the emission is detected by a collection lens and a fiber spectrophotometer (see Materials and Methods section) at an angle $\theta_{det}$ complementary to $\theta_{in}$. Figure 3b shows the PL spectrum of the d-MIM as a function of the detection angle $\theta_{det}$, where the sample was rotated with respect to the detection and pump paths. The dispersion of the emission peak with the detection angle is plotted in Figure 3c, and corresponds well with the angular behavior of the absorbance resonances of a d-MIM measured by ellipsometry (dashed blue and red lines).

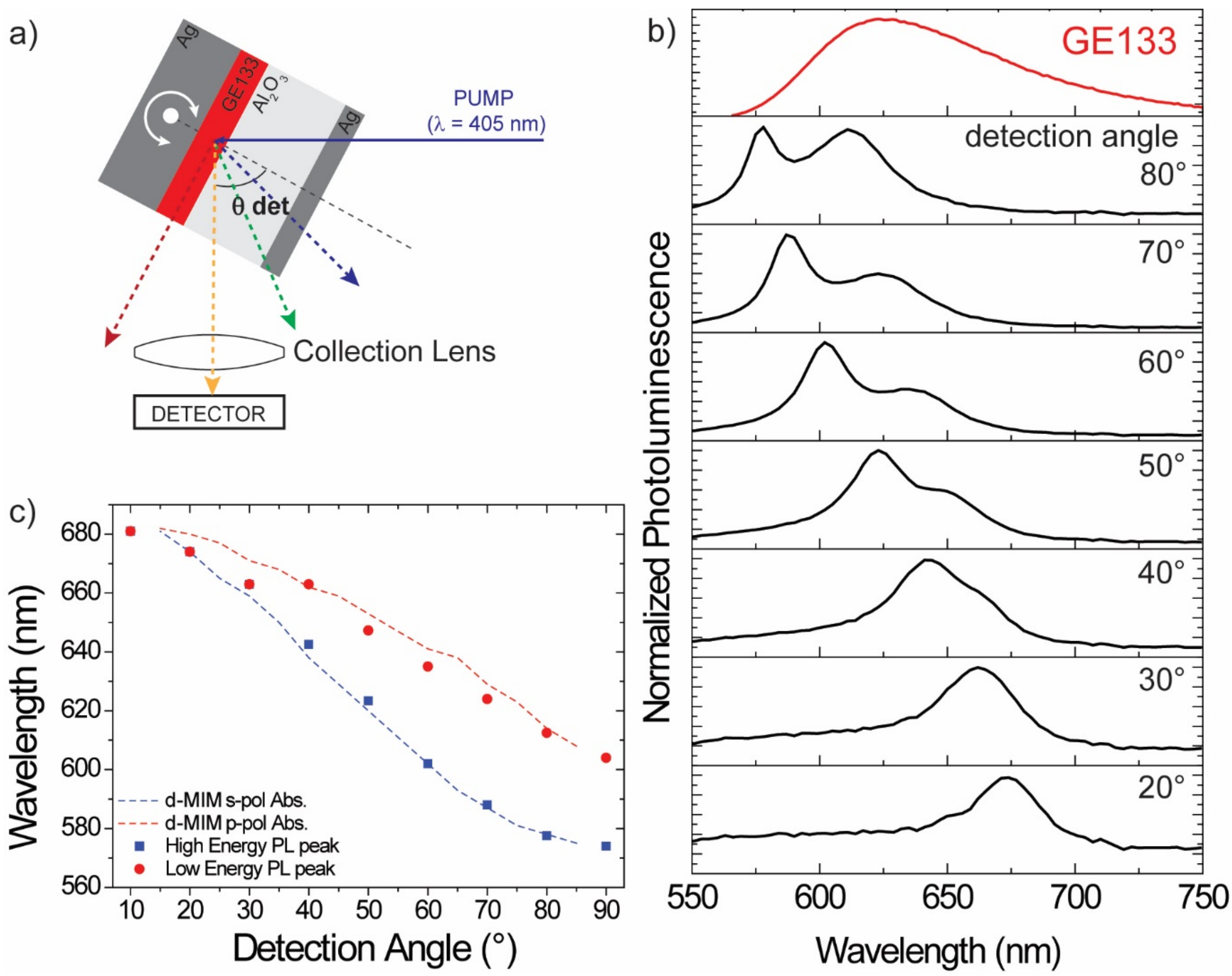

**Figure 3:** (a) Sketch of the experimental setup for the angle-dependent PL measurements. (b) PL as a function of the detection angle (black curves), and the emission of the bare GE133

(red curve). (c) Photoluminescence peak positions (markers) and experimentally measured p- and s-polarized absorbance modes of the d-MIM plotted *versus* the detection angle.

The radiation that is outcoupled from the nanocavity preserves the polarization of the p- and s-polarized resonances at the respective wavelengths. We confirmed this behavior with an analyzer in the detection path, as shown in **Figure 4a**. The PL was acquired for different analyzer angles, from 0° (p-polarization) to 90° (s-polarization), and the corresponding spectra are plotted in Figure 4b. At 0° only the low-energy p-polarized PL peak passes the analyzer, and at 90° only the s-polarized peak, confirming our polarization assignment. For intermediate angles, both p- and s- contribution pass the filter, with the expected gradual change in peak intensity from P to S with increasing angles.

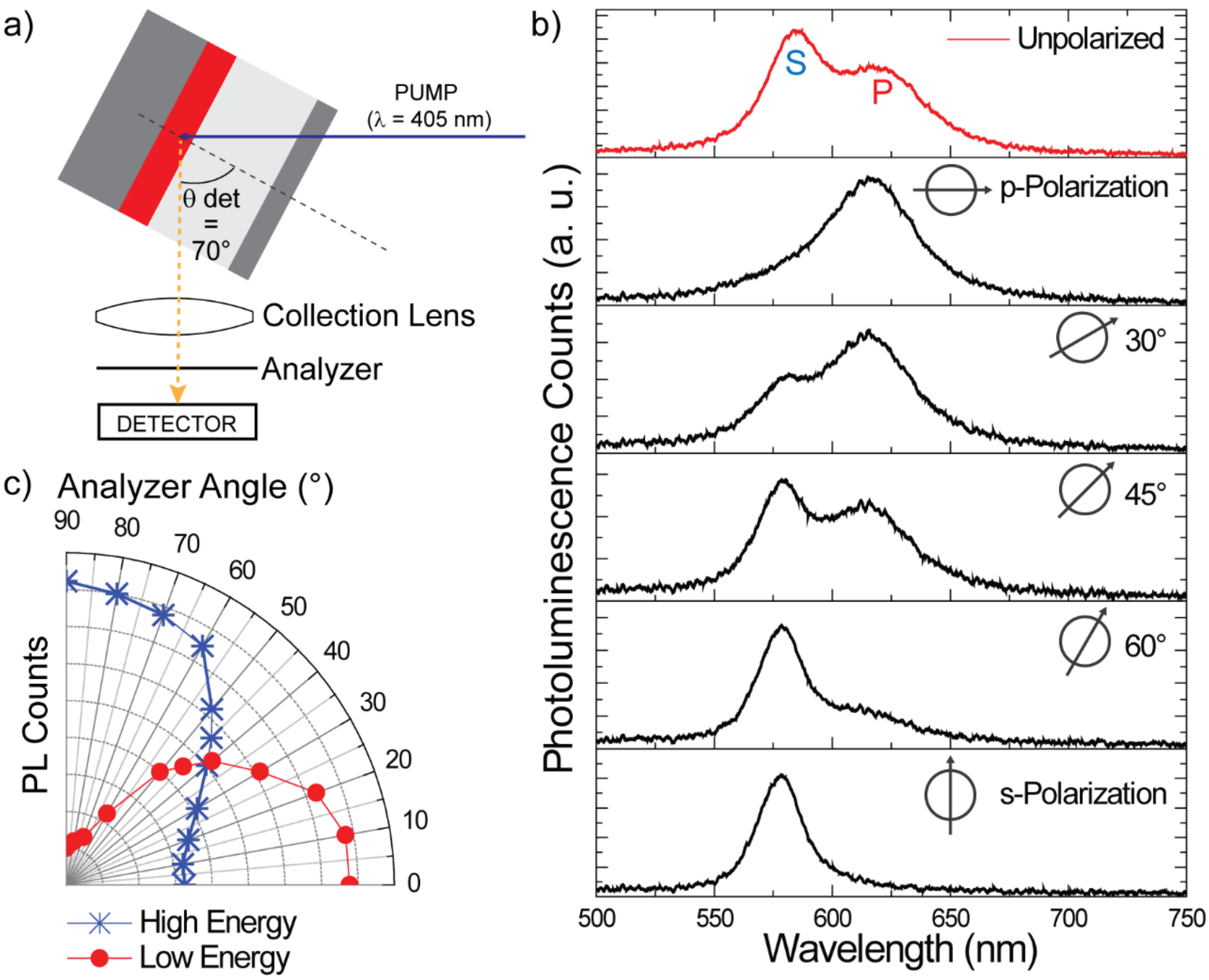

**Figure 4.** (a) Sketch of the experimental setup for the detection of the polarization of the emission. (b) Photoluminescence spectra acquired at different analyzer angles (black curves),

together with the not-analyzed emission spectrum of the d-MIM. (c) Polar analysis of the photoluminescence peaks as a function of analyzer angle.

The polar diagram of Figure 4c summarizes this behavior, showing that the radiated field manifests the typical profiles of the polarized emission. The intensity of the low-energy p-polarized mode is highest at 0° and lowest at 90°, while that of the s-polarized high-energy mode is complementary, with a maximum at 90° and minimum at 0°.

**3. Conclusion**

In conclusion, we investigated the capability of a fluorophore-doped MIM nanocavity to reshape the photoluminescence properties of the embedded emitter. We demonstrated both theoretically and experimentally that the outcoupled spontaneous emission of a gain material consisting of a film of push-pull chromophores (GE133) within the MIM cavity is determined by the angular dispersion of the d-MIM cavity. Moreover, the emission from the d-MIM cavity preserves the polarization state imposed by the cavity. d-MIM cavities with such angular and polarization selective radiative properties are highly appealing for applications in integrated optical devices and waveguiding. On one hand, such layered device structures can be prepared on wafer scale without the need of lithographic patterning. On the other hand, the discussed angular and polarization selective properties should not be limited to the optical far-field, and therefore could also unfold polarization wavelength division multiplexing (P-WDM) applications at the nanoscale. The tunability of the cavity resonances can be extended by using coupled MIM cavities that form a MIMIM system. In that case it is possible to tailor the cavity resonances both to the optical pump frequency and to the emission wavelength of the dye. Finally, d-MIM systems can also be envisioned on flexible substrates, which could pave the way towards more sophisticated geometries. For example, the angular dispersion of the flat cavity could be compensated *via* suitable curvature to obtain customizable near-field focused emission and nanolenses.

## 4. Experimental Section

*Fabrication and characterization of the d-MIM structures:*

We used a multistep procedure consisting of (i) Electron Beam (Kurt J. Lesker PVD 75) deposition of the Ag thick backreflector, (ii) spin coating of a solution 0.02 M of GE133 fluorophore in $CHCl_3$ (2000 rpm, 1 min) with control of the film thickness via profilometry (Veeco Detak 150) and Ellipsometry, (iii) Electron Beam deposition of $Al_2O_3$ thin films to tune the cavity resonance to the emission of the GE133, (iv) E-Beam deposition of the top 30 nm Ag layer to complete the cavity.

After every step, the optical properties of the sample were characterized ellipsometrically with a Vertical Vase ellipsometer by Woollam, in the range from 300-900 nm with a step of 3 nm. The angular dispersion of the d-MIM was measured via p- and s- polarized reflectance experiments in a range between 15° and 85° with a step of 5°.

Photoluminescence spectra were acquired with an Edinburgh Instruments spectrofluorometer (FLS920) with a spectral resolution of 2 nm, where the sample was illuminated at 480 nm with a Xe lamp coupled to a monochromator.

*Angular and polarization dependent photoluminescence setup:*

Angular dependent PL (ADPL) spectra were collected as sketched in panel (a) of Figure 3. A pulsed laser beam at 405 nm with pulse width ~100 fs and repetition rate of 1 MHz was focused on the sample with a piano-convex lens (F = 50 mm). The sample was mounted on a 3D rotating stage. ADPL spectra were acquired at angular steps of 5° by collimating the emitted light with two broad acceptance angle piano-convex lenses ($F_{total}$ = 70 mm) to a fiber spectrophotometer (Ocean Optics HR4000). Polarization emission experiments were carried out by using a linear polarizer (analyzer) in the detection path between the collection lenses and the fiber.

*COMSOL simulations:*

The emitter embedded in the cavity was modelled as a horizontally oriented point dipole placed at the center of the dielectric core of the d-MIM superabsorber. Refractive indices of the materials were directly measured from the deposited films via spectroscopic ellipsometry, As boundary conditions, appropriately swept Perfectly Matched Layers were adopted at the top, bottom, right and left sides of the model and continuity periodic conditions were fixed at the left and right side of the domain to model an infinite domain in the horizontal direction. The d-MIM cavity was meshed via a free triangular texture with a maximum element size of 2 nm.

**Supporting Information**

Supporting Information is available from the Wiley Online Library or from the author.

- Angular dependence of the Q-Factor of the p- and s-polarization modes
- COMSOL simulation of the angular dispersion of s-polarized modes
- Emission spectrum of a pristine film of GE-133
- Analysis of the d-MIM eigenmodes
- Decay lifetime of the pristine GE133, open cavity, and asymmetric and symmetric closed cavity d-MIM systems.
- Effective refractive index of the d-MIM

**Acknowledgements**

We thank Dr. Mario Miscuglio (George Washington University) for fruitful discussions on the COMSOL simulations. The research leading to these results has received funding from the European Union under the Marie Skłodowska-Curie RISE project COMPASS No. 691185, and from the HORIZON 2020 ATTRACT project “TEHRIS” ID n° 777222.

Received: ((will be filled in by the editorial staff))
Revised: ((will be filled in by the editorial staff))
Published online: ((will be filled in by the editorial staff))

The table of contents entry should be 50–60 words long and should be written in the present tense and impersonal style (i.e., avoid we). The text should be different from the abstract text.

**Keyword** Photonic cavities ((choose a broad keyword related to the research))

V. Caligiuri*, G. Biffi, M. Palei, B. Martin-Garcia, P. R. Devi, Y. Bretonnière, and R. Krahne*

**Angle and Polarization Selective Spontaneous Emission in Dye-doped Metal/Insulator/Metal Nanocavities**

ToC figure ((Please choose one size: 55 mm broad × 50 mm high **or** 110 mm broad × 20 mm high.  Please do not use any other dimensions))

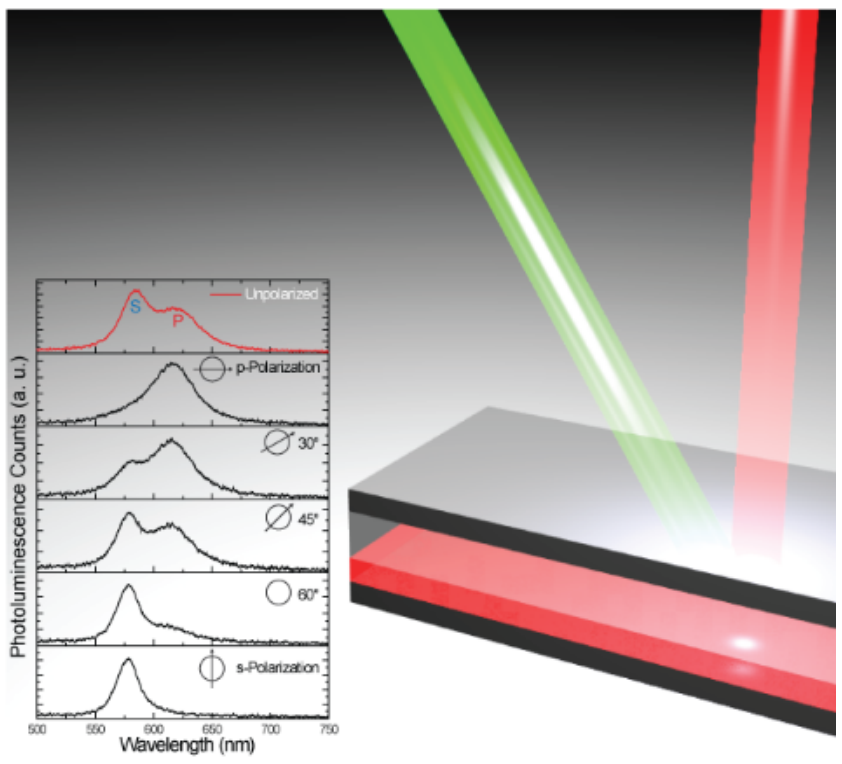

((Supporting Information can be included here using this template))

# Supporting Information

**Title** ((no stars))

*Author(s), and Corresponding Author(s)** ((write out full first and last names))

((Please insert your Supporting Information text/figures here. Please note: Supporting Display items, should be referred to as Figure S1, Equation S2, etc., in the main text…)